# Quantitative particle agglutination assay for point-of-care testing using mobile holographic imaging and deep learning

Yi Luo[1,2,3 †], Hyou-Arm Joung[1,2,3†], Sarah Esparza[3], Jingyou Rao[4], Omai Garner[5], Aydogan Ozcan[1,2,3 *]

[1]Electrical & Computer Engineering Department, [2]California NanoSystems Institute (CNSI), [3]Bioengineering Department, [4]Computer Science Department, [5]Department of Pathology and Laboratory Medicine, University of California, Los Angeles, California 90095, United States
*Corresponding author: *E-mail: ozcan@ucla.edu.

† These authors contributed equally.

**Abstract**: Particle agglutination assays are widely adapted immunological tests that are based on antigen-antibody interactions. Antibody-coated microscopic particles are mixed with a test sample that potentially contains the target antigen, as a result of which the particles form clusters, with a size that is a function of the antigen concentration and the reaction time. Here, we present a quantitative particle agglutination assay that combines mobile lens-free microscopy and deep learning for rapidly measuring the concentration of a target analyte; as its proof-of-concept, we demonstrate high-sensitivity C-reactive protein (hs-CRP) testing using human serum samples. A dual-channel capillary lateral flow device is designed to host the agglutination reaction using 4 µL of serum sample with a material cost of 1.79 cents per test. A mobile lens-free microscope records time-lapsed inline holograms of the lateral flow device, monitoring the agglutination process over 3 min. These captured holograms are processed, and at each frame the number and area of the particle clusters are automatically extracted and fed into shallow neural networks to predict the CRP concentration. 189 measurements using 88 unique patient serum samples were utilized to train, validate and blindly test our platform, which matched the corresponding ground truth concentrations in the hs-CRP range (0-10µg/mL) with an $R^2$ value of 0.912. This computational sensing platform was also able to successfully differentiate very high CRP concentrations (e.g., >10-500 µg/mL) from the hs-CRP range. This mobile, cost-effective and quantitative particle agglutination assay can be useful for various point-of-care sensing needs and global health related applications.

## Introduction

Particle agglutination assays are widely used immunological tests based on antigen-antibody interactions[1,2]. Latex particles are sensitized through the adsorption of antibodies onto their surfaces. Once the sample is introduced, the corresponding antigens attach to the antibody binding sites and the micro particles form clusters due to the target antigen's capability of binding to different sites simultaneously. The amount of agglutination between the particles is indicative of the amount of antigen present in a sample. Particle agglutination assays have been used to test for antigens in a number of bodily fluids, including e.g., saliva, urine, cerebrospinal fluid, and blood. A range of illnesses can be diagnosed using particle agglutination assays, including bacterial, fungal, parasitic and viral diseases[3–12]. Its major advantages in point-of-care diagnosis include short reaction time, low sample volume, low-cost, and high specificity. The operation of conventional particle agglutination assays includes two steps. First, an expert will mix the sample of interest with a pre-processed liquid that contains antibody-coated particles. The turbidity of the mixture then changes as agglutination occurs during the mixing process. Most often, an expert will compare the mixture to a pre-existing standard of turbidity to give an estimation of the antigen's concentration using the naked eye. One of the barriers to its wider application lies in the assay's low sensitivity and lack of quantitative measurements[13,14]. A simple practice to get semi-quantitative analysis is to perform multiple tests simultaneously with pre-diluted samples of different concentration gradients. Other research has also focused on measuring optical turbidity or light scattering with the help of a spectrometer to give quantitative readouts[15–21], which relatively complicates the system and consumes a large sample volume (e.g., 1 mL per test[15]). Another possible solution, rather than solely measuring the turbidity, is to use optical microscopes to monitor the agglutinated particle clusters in the assay. However, such an improvement will further complicate the diagnostic system and relatively increase the cost per test[22]. A wave of democratizing optical microscopes for point-of-care applications has been observed in the past decade[23–30]. Among these, lens-free holographic microscopes gained interest given their advantages, such as cost-effectiveness, portability, high resolution, and large field of view[31]. A lens-free holographic microscope uses a partially coherent light, usually from a light emitting diode (LED), to illuminate a thin and transparent sample that is placed right above a complementary metal oxide semiconductor (CMOS) sensor-array, by which an inline hologram of the sample is recorded[31,32]. Portable microscopy devices based on this method have been applied in many fields, including water quality monitoring[33,34], pollen detection[35,36], and virus sensing[37–40]. By combining holographic microscopic imaging with deep learning, challenging tasks can be achieved including e.g., birefringent crystal detection[41] and virtual staining[42].

In this paper, we demonstrate a rapid and cost-effective quantitative particle agglutination assay using deep learning-based analysis (Fig 1(a)), automatically measuring high sensitivity C-reactive protein (hs-CRP) levels in human serum samples. CRP is a general biomarker produced by the liver as a response to inflammation in the body which has a concentration in the range of up to ~1000μg/mL, whereas hs-CRP, in the range of 0.5 to 10μg/mL[43], is an indicator for the risk of myocardial dysfunction and heart failure. In our particle agglutination assay, human serum samples with various CRP concentrations are mixed with latex particles and the mixture is immediately loaded into a custom-designed, dual-channel capillary lateral flow device (Fig. 1(b)). Agglutination takes place automatically while the mixture flows through the capillary channels of the lateral flow device (Fig. 1(c)) for ~3 min without any further operation steps. Time-lapsed holographic images of the mixture are acquired in real time with a mobile lens-free microscope (Figs. 1(d) and (e)) to monitor and quantify the agglutination. Two neural networks are designed and trained to work sequentially to measure the concentration of hs-CRP and differentiate it from acute inflammation (>10 - 1000μg/mL). For this, a classification network is first used to identify CRP concentrations higher than 10μg/mL and a quantification network is sequentially applied specifically to hs-CRP range to predict the concentration (≤ 10μg/mL) of the sample. To demonstrate the success of this platform, we measured 88 patient serum samples, 65 for training and validation and 23 for blind testing, covering a CRP concentration range of 0.2 to 500 μg/mL, and achieved an $R^2$ value of 0.912 on blind testing set with respect to the ground truth measurements, captured by an FDA-approved clinical instrument. We foresee that this computational sensing platform can be used in point of care settings to provide rapid and cost-effective measurements of various analytes.

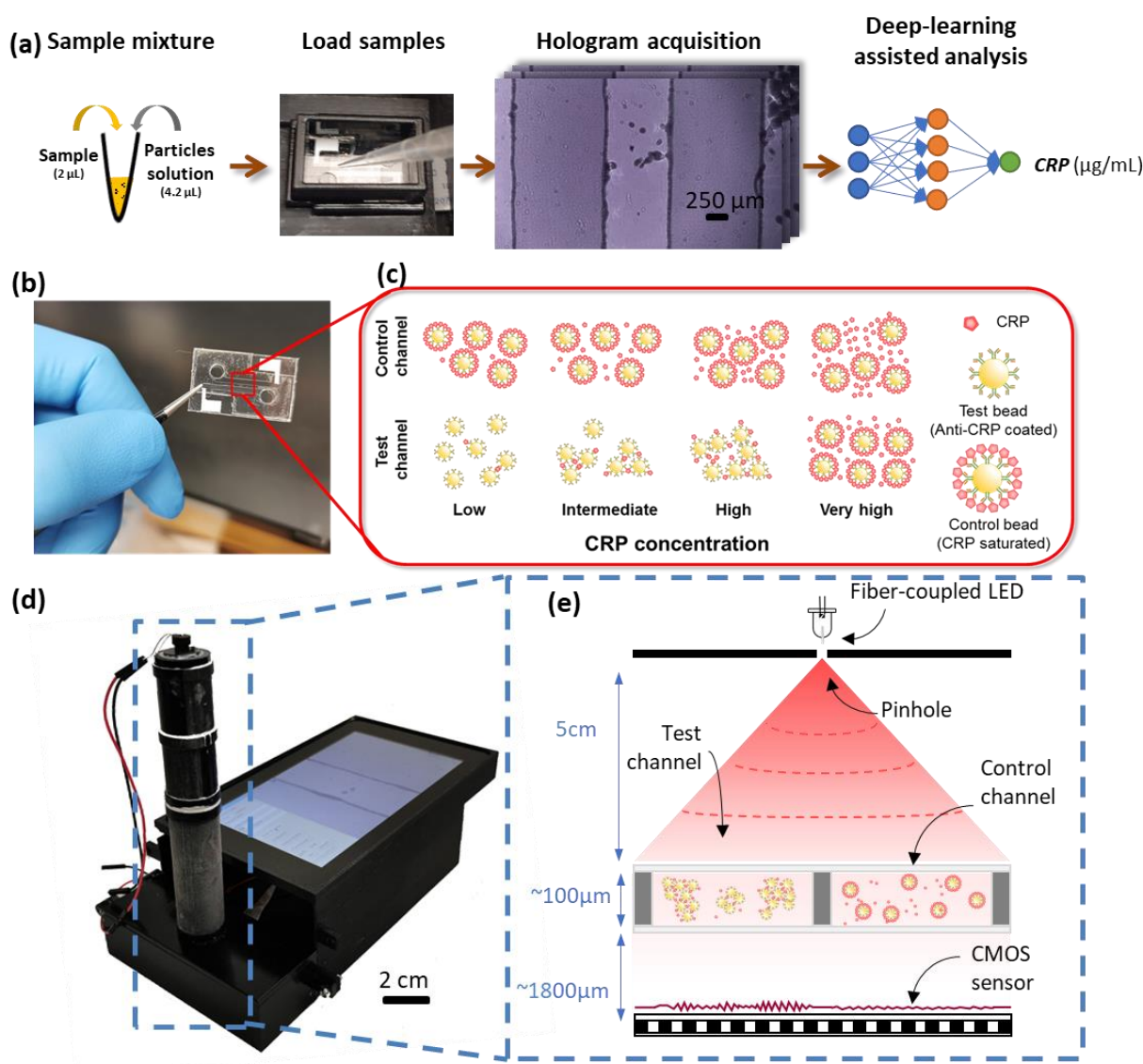

**Fig. 1.** Quantitative particle agglutination assay using dual-channel capillary-based lateral flow device and mobile lens-free microscopy. (a) Operation procedure of the quantitative particle agglutination assay. (b) A photograph of the dual-channel capillary lateral flow device. (c) Schematic of particle agglutination inside both the test and control channels under different CRP concentration levels. (d) A photograph of the mobile lens-free microscopy device. (e) Schematic drawing of the system.

## Materials and Methods

**Dual-channel capillary lateral flow device fabrication**

The dual-channel capillary lateral flow device is composed of different types of sheet materials (Supplementary Fig. S1). Optically clear transparent sheets (AZ42, Aztek Inc., Irvine, CA, USA) are cut into different shapes to serve as the floor and ceiling of the device using a laser cutter (60W Speedy 100 CO2 laser, Trotec, USA). A double-sided tape (12X12-6-467MP, 3M Inc, USA) is cut to form the side walls of both the test and control channels, as well as the loading posts. The absorbing membranes are cut from MMM 0.8 sheet (T9EXPPA0800S00M, Pall Corporation, USA). Before the assembly, the transparent sheets are cleaned using an ultrasonic bath. Then, the transparent floor piece, double sided tape, absorbing pads and transparent ceiling piece are stacked together to form a sandwich-like structure.

**Preparation of test and control particles**

Both the control and test particle solutions are prepared using the CRP Latex Reagent component of the CRP Latex Test Kit (310-100, Cortezdiagnostics Inc, USA), which has an average particle diameter of 0.81 µm. To prepare the test particles, 100µL of Latex Reagent is centrifuged at 3000 rpm for 10 min. After removal of the resulting supernatant, the beads are re-suspended in an equal volume of glycine buffer. The CRP saturated *control* particles are prepared by saturating antibody binding sites on the test particles; for this, the particles are diluted three times by PBS buffer and CRP antigen (30-AC10, Fitzgerald) is added to the solution to reach a final CRP concentration of 0.5mg/mL. Following a 2-hour incubation with an orbital shaker and the addition of 1% BSA, the prepared particles are stored at 4 °C.

**Assay procedures**

To perform the assay, we add 5 µL of the activation buffer (0.5% tween 20 in DI water) into both test and control channel inlets. We then dry off the channels and place the sensor onto the CMOS image sensor with a custom-designed holder. Next we mix 2 µL of the serum sample with 4.2µL of test and control particle solutions individually, and load them into the corresponding inlets. Following this, the measurements start, recording the in-line holograms of the channels for 3 min.

**Collection of clinical samples**

The use of human serum samples was approved by UCLA IRB (#19-000172) for CRP testing. The CRP levels of these patient samples were measured by CardioPhase hsCRP Flex® reagent cartridge (Cat. No. K7046, Siemens) and Dimension Vista System (Siemens) at UCLA Health System, which constituted our ground truth measurements.

**Mobile lens-free microscope**

A mobile lens-free microscope was developed for monitoring of the particle agglutination assay reactions inside the capillary device. A fiber-coupled light emitting diode (LED, peak wavelength: 850 nm) is used to illuminate the capillary device to form inline-holograms[31]. A CMOS sensor (IMX 219, Sony Inc.) is placed right beneath the sample holder (with a sample-to-sensor distance of ~ 2.5mm) to capture the holograms at a frame rate of 1 fps. The illumination LED and the CMOS sensor-array are controlled by a Raspberry Pi microcomputer with a customized graphic user interface (GUI) that is programmed using Python.

**Particle localization measurements using multi-height digital back propagation**

The captured time-lapsed holograms are automatically processed using MATLAB. The test and control channels are first identified and cropped out. The background of each channel, which contains randomly located dust particles that are attached to the bottom of the microfluidic chip, is estimated using the first five frames of the image sequence. The axial distance ($z_2$) between the microfluidic chip and the CMOS sensor is approximated by auto-focusing on these immobile particles[44]. Knowing the thickness of the transparent sheet (100µm) and the height of the channels (100µm), the raw hologram of each channel is back propagated to multiple axial locations using the angular spectrum method[40,45], ranging from $z_2$ + 100µm to $z_2$ + 200µm with an axial step size of 10µm. At each axial plane, a binary particle map is generated using the amplitude channel of the propagated hologram, by applying a threshold (i.e., mean amplitude value minus three standard deviations). The binary masks of ten different axial locations are summed up after removing false detections that only appeared in a single axial location. Immobile/stationary particles were also removed by comparing two consecutive frames of the binary masks. Morphological analysis is applied to this merged binary mask, which results in the estimation of the total particle area ($A_t(t)$ and $A_c(t)$) and the total particle number ($n_t(t)$ and $n_c(t)$), as a function of time ($t$), in the test and control channels, respectively.

**Neural network architecture**

Two shallow neural networks (see Supplementary Fig. S2) were trained to quantitatively measure the CRP concentration over a large dynamic range. The total particle area and the total particle number in both the test and control channels are time-averaged with a non-overlapping sliding time-window and used as inputs to train the neural networks. The classification network (Supplementary Fig. S2a) contains one fully-connected hidden layer with 2048 neurons and one output layer with 2 neurons, ($O_1, O_2$), which are used to indicate if the CRP concentration is below or above 10 μg/mL. Its input vector is composed of time-averaged $A_t(t)$, $A_c(t)$, $n_t(t)$ and $n_c(t)$ with a window size of 1, 1, 5 and 10, respectively. A cross-entropy loss function ($L_1$) is used to calculate the error gradients used in the training phase:

$$L_1 = \frac{1}{N}\sum_{i=1}^{N} -(y_i \log(p_i) + (1 - y_i) \log(1 - p_i)) \qquad (1)$$

where $y_i$ is a binary indicator (the ground truth label), representing if the measured CRP concentration is above 10 μg/mL or not, for each measurement $i$, in a training batch of $N$ different measurements. $p_i$ indicates the probability whether the CRP concentration is higher than 10μg/mL or not for a given measurement $i$. It is calculated using the output values of the network $\boldsymbol{O} = [O_1, O_2]$ as

$$p = \frac{\exp(O_1)}{\exp(O_1) + \exp(O_2)} \qquad (2)$$

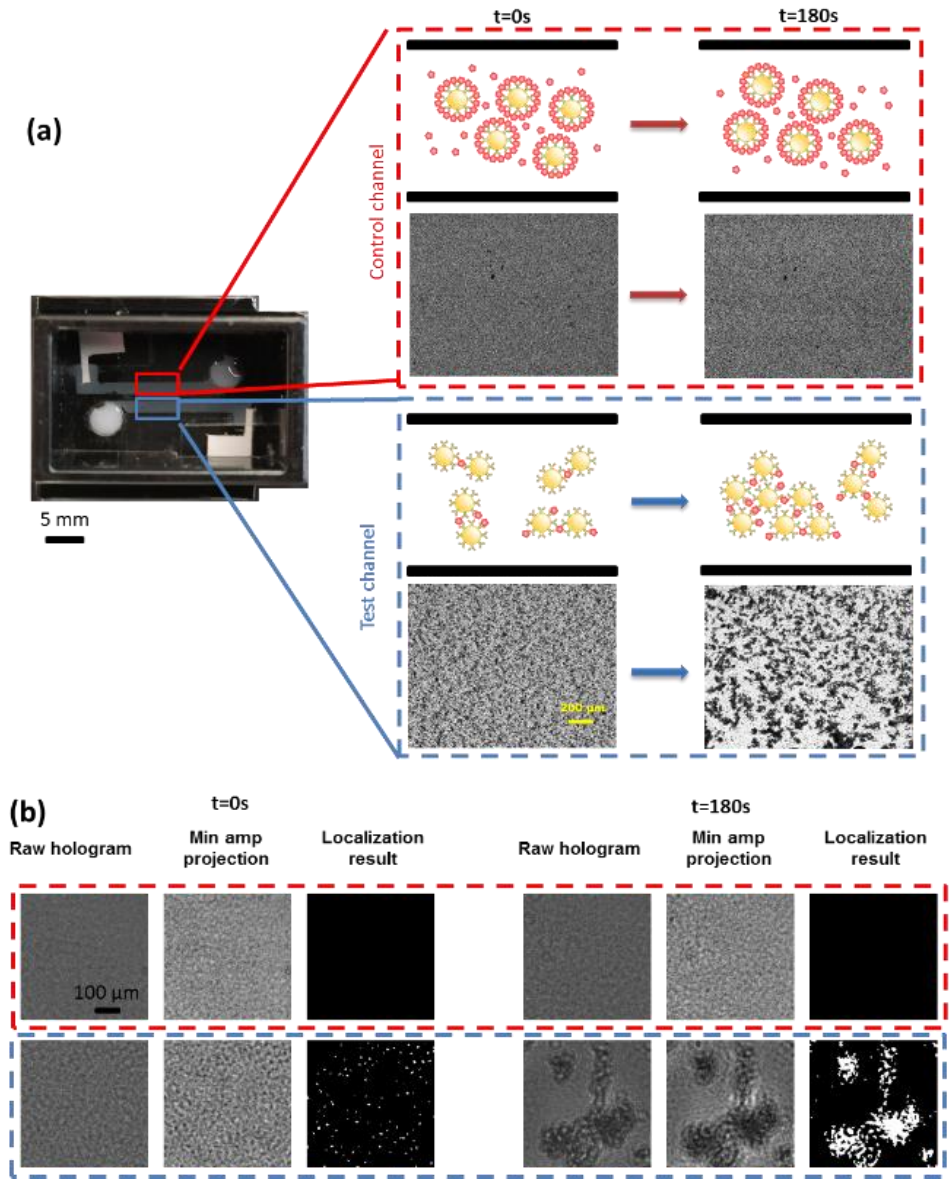

**Fig. 2.** Operation principle of the quantitative particle agglutination assay. (a) Schematic drawing and lens-free microscopic images of particle clusters inside both the control and test channels at different time points. (b) Inline holograms of the control and test channels at different time points. The processed minimum amplitude projection and particle localization results are also shown for each channel.

The quantification network (Supplementary Fig. S2b), on the other hand, contains two fully-connected hidden layers with 32 and 8 neurons separately, and one output layer with a single neuron ($Q$), outputting the predicted CRP concentration within the hs-CRP range. The time-averaging window sizes for $A_t(t)$, $A_c(t)$, $n_t(t)$ and $n_c(t)$ are equal to 30, 30, 1 and 30, respectively. A mean-square-error loss function ($L_2$) is applied for the training of the quantification network, defined as:

$$L_2 = \frac{1}{N} \sum_{i=1}^{N} \sqrt{(Q_i - C_i)^2} \qquad (3)$$

In Eq. (3), $Q_i$ is the value of the single output neuron, representing the predicted CRP concentration, and $C_i$ is the ground truth concentration measured by the gold standard instrument for each measurement $i$.

The hyper-parameters of both neural networks (e.g., the number of neurons and sliding window sizes for $A_t(t)$, $A_c(t)$, $n_t(t)$ and $n_c(t)$) are optimized through a greedy search. For each parameter search, the corresponding neural network was trained for 5 times with 500 epochs in each training. The validation loss was then averaged to find the best candidates. After optimizing all the hyper-parameters, both networks were trained using the Adam optimizer for 1,000 epochs. At the beginning, the learning rate was set to $10^{-4}$. The validation loss was calculated after every epoch of training and a learning rate scheduler was adopted to monitor the validation loss so that the learning rate was reduced by a factor of two if there was no improvement in 100 consecutive epochs of training. The networks were composed using Pytorch and trained on a desktop computer (Origin PC Corp., FL, US) using a CPU only. The typical training time for classification and quantification networks is ~30 sec and ~60 sec, respectively. For blind inference, the classification and quantification neural networks on average took less than 0.1ms per test using a desktop computer with 64 GB memory and i9-7900X CPU (Intel corp., CA, US).

## Results

**Quantitative particle agglutination assay and portable holographic reader**

A lateral flow particle agglutination assay was developed to quantitatively measure the CRP concentration of serum samples by monitoring the particle agglutination reaction between CRP and antibody coated latex particles. The assay was composed of a custom-designed, low-cost dual-channel capillary lateral flow device to host the antigen-antibody interactions (Figs. 1(b) and (c)) and a mobile lens-free microscope to monitor and quantify the agglutination process (Figs. 1(d) and (e)). The operation principles of the lateral flow particle agglutination assay are depicted in Fig 2. Two microliters of human serum sample under test is mixed with test beads (antibody-coated latex micro-particles) and control beads (CRP saturated particles) separately (see the Materials and Methods section for details), i.e. 4µL of serum sample is consumed per test. Without any incubation, the mixtures are directly loaded into the test and control channels of the dual-channel capillary lateral flow device. The device is custom-designed and fabricated using low-cost sheet materials, with a total material cost of 1.79 cents per chip, which can be further reduced through mass-production and economies of scale (Supplementary Fig. S1).

A sheet-tape-sheet sandwich structure was manually assembled to form the test and control channels. An absorption membrane was inserted at the outlet of each channel. Water evaporation on the membrane provides the driving force for continuous flow. The diffusion of both CRP and latex particles in the laminar flow enabled the antigen-antibody reaction in the test channel, resulting in agglutinated particle clusters with their size varying as a function of the test time and the CRP concentration in serum (Fig 2(a)). In the control channel, however, given that the antibodies have already been saturated with CRP (by design), only non-specific agglutination between latex particles and unknown proteins in serum can happen (Supplementary Fig. S3)[46]. The incorporation of the control channel in our sensor design provided us a self-calibration tool for mitigating potential false positive agglutination.

During the total assay time (3 min), time-lapsed inline holograms were acquired using a mobile lens-free microscope at 1 frame/sec. The captured holograms (Fig 2(b)) were automatically processed and for each frame, the locations of the test and control channels were digitally cropped out. Given the existence of a large number of particles whose diameter is close to the illumination wavelength, the liquid mixtures in both channels are highly scattering. To localize and measure the forming particle clusters, a multi-height back-propagation algorithm was applied to each channel and the total particle number as well as the total particle area were extracted (Fig. 3(a); see the Materials and Methods section for details). Time series of these features were utilized to determine the CRP concentration of the serum sample using two sequentially collaborating, trained neural networks (Fig. 3(b)), i.e., a classification network and a concentration quantification network, respectively. Both of these neural networks are shallow with a low number of trainable parameters, to ensure rapid inference speed and avoid overfitting (see the Materials and Methods section and Supplementary Fig. S2).

**Quantification of CRP concentration using deep learning**

88 human serum samples were collected from different patients with various CRP concentrations. 144 different measurements were conducted on these clinical samples (duplicate measurements were conducted on 56 samples). Given that only three out of 88 serum samples had CRP concentrations higher than 10μg/mL, additional acute inflammation samples were created by spiking CRP into clinical samples to achieve a concentration of >10μg/mL. For this purpose, three clinical samples with original CRP concentrations lower than 0.2μg/mL were spiked to achieve five different CRP concentrations (20, 50, 100, 200 and 500μg/mL), forming 15 additional samples to represent a concentration range of >10μg/mL. A total of 45 measurements were performed on these CRP-spiked, additional samples (triplicate measurements on each sample). Therefore, the total number of CRP measurements that we have made with our sensor platform is 144 + 45 = 189.

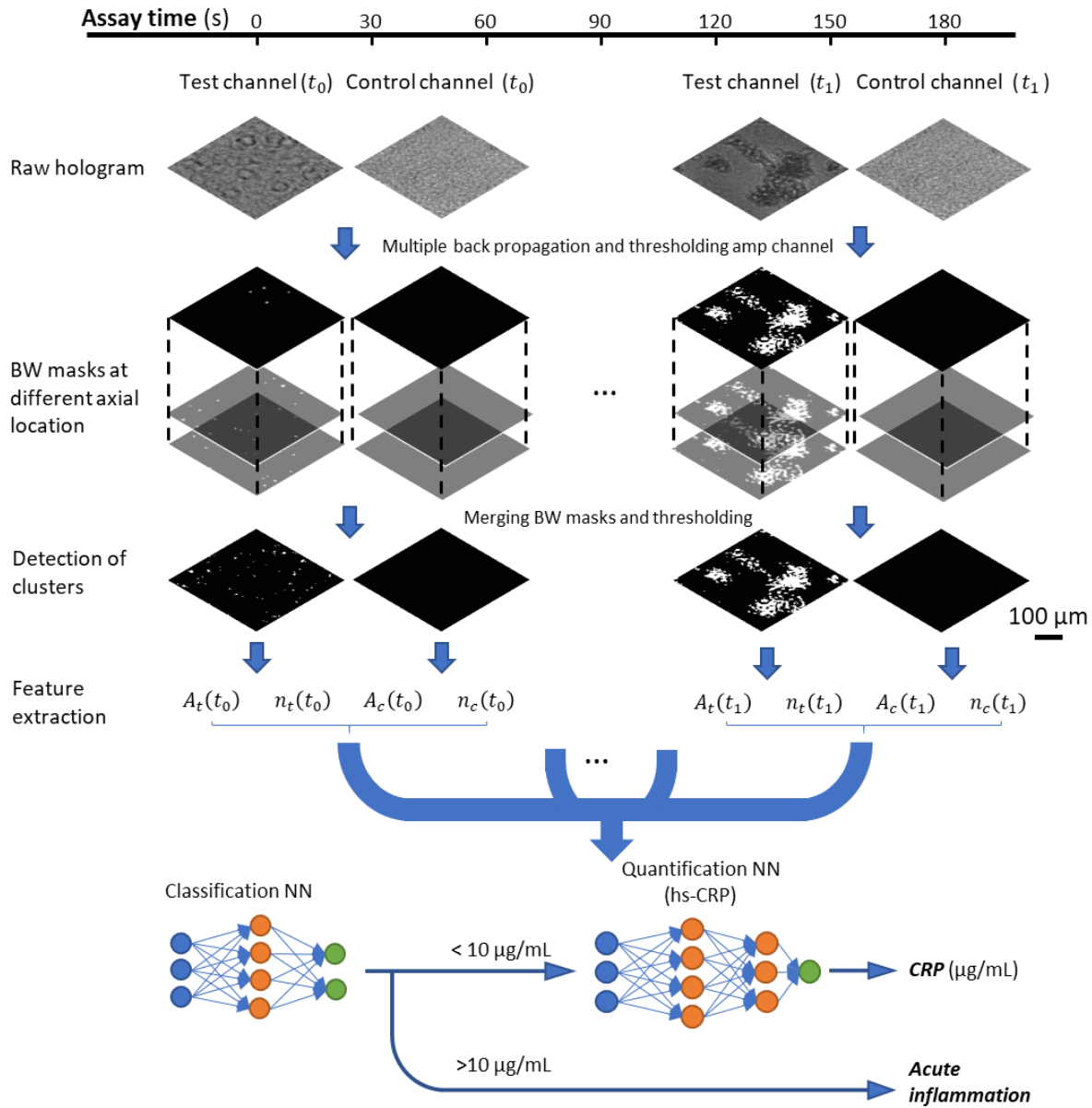

**Fig. 3**. Image processing pipeline. For each holographic frame, the test and control channels are automatically cropped out. A multi-height digital back-propagation algorithm is applied to both channels. The extracted features are input into two shallow neural networks to determine the CRP concentration of the serum sample.

Conventional particle agglutination assays suffer from false-negative diagnosis of high concentration cases due to the saturation of antibody's binding sites by excessive antigens, also as known as the hook effect[47]. The impact of the hook effect can also be clearly seen in our raw measurements. The total cluster size measured at the end of the assay time in the test channel of the serum samples with different CRP concentrations is illustrated in Fig 4(a). The cluster size increased with increasing concentration of CRP, until it reached a maximum at around 10μg/mL. Further increases of CRP concentration saturated the antibody's binding sites and reduced the cluster size (Fig 4(a)). To overcome the hook effect and accurately identify the unknown CRP concentration over a large dynamic range of CRP concentrations, two different neural networks were designed to work sequentially (Fig 3(b)). The first one, i.e., the classification network, was designed to distinguish if a sample has a CRP concentration higher than 10µg/mL or not. The network was trained (and validated) using 96 (49) different measurements out of all 189 measurements, covering CRP concentrations ranging from 0 to 500µg/mL. 44 measurements from 23 different patients that were not used in the training and validation datasets were used to form the testing dataset (see Materials and Methods section for details). The second network, i.e., the quantification network, was designed to quantitatively determine the hs-CRP concentration in the sample (covering 0-10µg/mL). The network was trained using the same measurement dataset separation reported earlier, but eliminating all the measurements on samples with a CRP concentration that is higher than 10µg/mL. As a result, the training, validation and testing sets of the quantification network contained 71, 33 and 31 different measurements, respectively.

The decision-making performance of this two-network based computational sensing system is depicted in Fig. 4(b). For hs-CRP, the predicted concentration by our device is plotted with respect to the corresponding ground truth concentration measured for each sample at UCLA Health System. The dashed line indicates a perfect prediction, i.e., $y = x$. For samples with high CRP concentration (>10µg/mL), the confidence level of the classification result is also presented in the same plot. These results reveal that the classification network successfully separated all the samples in the blind testing dataset based on their concentration (acute inflammation vs. hs-CRP), overcoming the hook effect[47]. With 31 blind tests quantified in the hs-CRP range, the $R^2$ value was found to be 0.912 with respect to the $y = x$ line, demonstrating the inference success of the quantification neural network.

To further highlight the capabilities of this neural network-based inference of the target analyte concentration, in Fig. 4 we report our measurement results (duplicate measurements marked with green dots) on a clinical sample with 0.7µg/mL of CRP measured by the gold standard instrument. At the end of the assay time (180 sec), the total particle area in the test channel provided a strong false-positive signal (green dots in Fig. 4(a)), which can occur frequently in clinical testing due to unknown proteins present in patient serum[2]. Even for these challenging tests that would normally result in a false positive, our quantification neural network successfully inferred the CRP concentration in these duplicate measurements as indicated with the green dots in Fig. 4(b), avoiding a false cardiovascular risk factor classification, which exemplifies the inference power of our imaging-based particle agglutination assay.

## Discussions

In this work, the antibody-antigen interaction is assisted by the laminar flow inside the dual-channel capillary lateral flow device. The flow rate is a key parameter to guarantee the stable reactivity of the assay. The size of the absorption membrane and the external humidity are critical factors in determining the flow rate inside the capillary channel. The membrane size we used in this work was optimized by evaluating the assay's reactivity under different flow rates and different humidity conditions (see Supplementary Fig. S4). Through these optimization experiments, we selected a membrane of size of 16mm² which resulted in an average flow rate of 1.45±0.3µL/min. In addition, we confirmed that the selected membrane operates stably at all the tested humidity levels (10-50%) except for 70%, through the corresponding comparisons of the reactivity and humidity evaluation results.

Several studies were reported in the literature to overcome the hook effect in sensor response by using advanced assay designs[48–51]. Compared to conventional particle agglutination assays, our platform provides kinetic information of the agglutination process, which is essential in overcoming the hook effect. Although being similar at the end of the entire assay time, the total particle/cluster area in the test channel for serum samples with low and very high CRP concentrations present different dynamic patterns as a function of time, tracked with our time-lapse holographic imaging system (Supplementary Fig. S5). Agglutination gradually occurred in low CRP samples, showing a slow increase in particle cluster area as a function of time. On the other hand, very high concentration CRP samples quickly saturated the binding sites of the antibodies on the latex particles in the first few seconds of the assay, and the imaged particle area stayed stable in the remaining assay time.

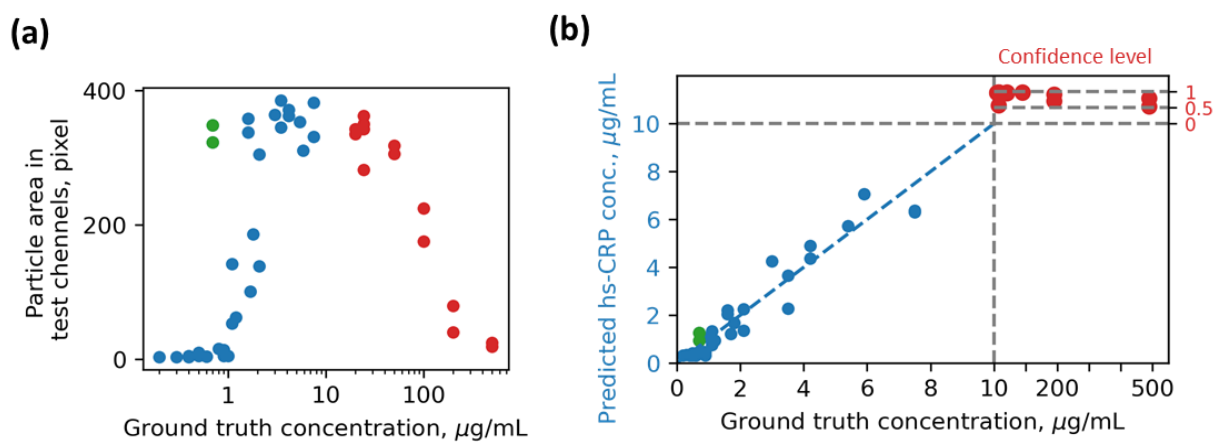

**Fig. 4**. (a) Particle area in the test channels measured at the end of the assay time (3 min) for different CRP concentration levels. (b) Neural network prediction of the CRP concentration inside the test channel.

In terms of digital processing of these spatio-temporal changes within the test channel, the particle localization algorithm that we employed significantly simplified the neural network structure. Although the raw acquired holograms were noisy, after the particles were localized using the multi-height back-propagation of each hologram, a shallow neural network architecture with a small number of neurons and trainable parameters was sufficient to quantify and classify the CRP concentration of serum samples over a large dynamic range. This shallow network architecture also shortened the inference time through each one of the networks: on average it took less than 0.1 ms per CRP test to have an output from the classification and quantification neural networks. With batch processing of multiple tests in parallel, this inference time can be further reduced.

## Conclusions

We demonstrated a rapid, simple, and cost-effective particle agglutination assay for point-of-care testing by using a custom-designed capillary lateral flow device and a mobile lens-free microscope. The agglutination of particles was captured, as a function of time, by a mobile lens-free microscope and digitally processed by two different neural networks for classification and quantification of the CRP concentration of the serum sample under test. This deep learning-assisted sensor has a low material cost (1.79 cents/test) and requires a small sample volume (4μL of serum per test), presenting a promising platform for various point-of-care sensing applications.